\documentclass[12pt]{iopart}
\usepackage{iopams}
\usepackage{amscd}
\usepackage{subeqn}
\usepackage{bm}
\usepackage[T1]{fontenc}
\usepackage[cp1250]{inputenc}
\renewcommand{\c}{\cdot}
\newcommand{\nn}{\nonumber}
\newcommand{\bea}{\begin{eqnarray}}
\newcommand{\eea}{\end{eqnarray}}
\newcommand{\ba}{\begin{array}}
\newcommand{\ea}{\end{array}}

\newcommand{\C}{{{\mathbb C}}}

\renewcommand{\bs}{\begin{subequations}}
\renewcommand{\es}{\end{subequations}}
\newcommand{\f}{\varphi}
\newcommand{\F}{\Phi}
\newcommand{\al}{\alpha}

\def\be{\begin{equation}}
\def\ee{\end{equation}}
\def\bn{\begin{enumerate}}
\def\en{\end{enumerate}}

\def\zd{{z^\dagger}}

\def\cpn{{\C P^{N-1}}}
\def\cp1{{\C P^1}}
\def\bp{{\bar{\partial}}}
\def\p{{\partial}}
\def\bxi{\bar{\xi}}

\def\bw{{\bar{w}}}

\newcommand{\qed}{\begin{flushright} $\square$
                  \end{flushright}
}
\begin{document}
\title[Painlev\'e property of $\cpn$ sigma models]{The Painlev\'e property of $\cpn$ sigma models}
\author{P P Goldstein$^{1}$ and A M Grundland,$^{2,3}$  }
\address{$^1$ Theoretical Physics Division, National Centre for Nuclear Research, Hoza 69, 00-681 Warsaw, Poland }
\address{$^2$ Department of Mathematics and Computer Sciences, Universit\'e du Quebec, Trois-Rivi\`eres. CP500 (QC) G9A 5H7, Canada }
\address{$^3$ Centre de Recherches Math\'ematiques. Universit\'e de Montr\'eal. Montr\'eal CP6128 (QC) H3C 3J7, Canada}

\ead{Piotr.Goldstein@fuw.edu.pl, grundlan@crm.umontreal.ca}
\begin{abstract}
We test the $\cpn$ sigma models for the Painlev\'e property. While
the construction of finite action solutions ensures their
meromorphicity, the general case requires testing. The test is
performed for the equations in the homogeneous variables, with
their first component normalised to one. No constraints are
imposed on the dimensionality of the model or the values of the initial
exponents. This makes the test nontrivial, as the number of
equations and dependent variables are indefinite. A $\cpn$ system
proves to have a $(4N-5)$-parameter family of solutions whose
movable singularities are only poles, while the order of the
investigated system is $4N-4$. The remaining degree of freedom,
connected with an extra negative resonance, may correspond to a
branching movable essential singularity. An example of such a
solution is provided.

\end{abstract}
 \vspace{2ex}
\begin{small}
\end{small}

\section{Introduction}

Numerous physical applications of models with effective
Lagrangians,  in particular the $\cpn$ sigma models
\cite{Bab1,Bab2,Dav,Landolfi,Manton,Raj,Zak}, made these models an
interesting subject to study \cite{Din,DZ,DHZ}. The question of
the integrability of the equations governing these models has
found an apparently positive answer in the works of Din and
Zakrzewski \cite{DZ}. Moreover, the linear spectral problem is
known for them, so (in principle) the initial problem may be
solved by the inverse scattering method. However the above results
only concern systems with finite action. On the other hand, if we
are interested in the dynamics of the systems, we start from the
corresponding Euler-Lagrange (EL) equations, which allow for a
much larger class of solutions. A natural question arises, as to
whether the equations remain integrable if we lift the assumption
of finite action. In the present paper we will discuss this
question and provide a self-contained approach to the subject.

The first approach which we try when testing a system of equations
for integrability is usually the Painlev\'e test in the form
introduced in \cite{ARS}, or its generalisation to partial
differential equations (PDEs) \cite{WTC}, with possible further
refinements (as discussed in \cite{Conte,CM,Mus}, which provide a
comprehensive review of the method).

In our case, the Painlev\'e test entails extra difficulties due to
the fact that the dimensionality of the $\cpn$ model and the
number of equations are arbitrary. Nevertheless, the test can be
carried out (Section 3).

In what follows, Section 2 contains a short summary of $\cpn$
models and possible methods of their description. We conclude that
section by selecting the description (system of PDEs) suitable
for the Painlev\'e test. In Section 3 we perform the test,
obtaining a `nearly general' local solution in the form of a
Laurent series. By `nearly general' we mean that our solution
provides $4N-5$ first integrals out of the total of $4(N-1)$ such integrals in the general
solution, i.e. it yields one integral fewer than the order of the system.
Section 4 contains a discussion of the missing first integrals. A
counterexample, i.e. an example of the non-Painlev\'e behaviour,
is given in the form of a solution which has an essential singular
manifold with branching. The manifold depends on four parameters
(although not on an arbitrary function), which means that the
position of the singularity depends on the initial conditions.

\section{$\cpn$ sigma models}
Sigma models describe complex systems by a simple Lagrangian
defined in terms of an effective field which lies in an
appropriate space, while the complexity remains in the metric of
the space.
\be\label{sigma}
\mathcal{L} =\sum_{i,j=0}^\infty g_{ij}dz_i d\bar{z}_j,
\ee
where $z_i,~\bar{z}_j$ represent the field variables in
$\mathbb{C}^N$, while $g_{ij}$ is the metric tensor. A bar over a
symbol denotes its complex conjugate (c.c.).

 The models prove to be rich in interesting
properties provided that the metric depends on the fields, i.e.
the model is nonlinear. Even simple nonlinear cases, like the
$\cpn$ models have many applications, from two dimensional gravity
to biological membranes \cite{Carrol, Gross,Landolfi}. In these
models the independent variables $\xi^1,\xi^2$ are from the
Riemann sphere or a 2D Minkowski space, $z\in S^N$, while the
differential in (\ref{sigma}) is expressed in terms of the
$z$-dependent covariant derivatives $D_{\mu}$ by
\be\label{cov}
D_{\mu}z=\partial_{\mu}z-(z^{\dagger}\cdot\partial_{\mu}z)z,\qquad\partial_{\mu}
=\partial_{\xi^{\mu}},\qquad\mu=1,2.
\ee
producing a Lagrangian density of the form
\be\label{lagr1}
\mathcal{L} = \frac{1}{4}(D_{\mu}z)^{\dagger}\cdot(D_{\mu}z),
\label{lagr-z}
\ee
where the convention of summation over repeating Greek indices is
assumed, $z$ are complex unit vectors in $\mathbb{C}^N$, a dagger
denotes the Hermitian conjugate, and $\p$ and $\bp$ are the
derivatives with respect to $\xi=\xi^1+ i\xi^2$ and $\bxi=\xi^1-
i\xi^2$ respectively. The normalisation of $z$ requires that
\be
\label{normalisation} z^\dagger\c z=1,\quad z=(z_0,...,z_{N-1}).
\ee
The EL equations corresponding to the Lagrangian (\ref{lagr1})
\be
D_{\mu}D_{\mu}z+(D_{\mu}z)^{\dagger}\cdot(D_{\mu}z)z=0,
\ee
are simple, but they are not suitable for testing the Painlev\'e
property: due to the normalisation (\ref{normalisation}), a pole
of $z$ has to correspond to a zero of $\zd$, at least for real
$\xi^\mu$. For the same reason, we do not analyse even simpler
equations satisfied by the rank-1 projectors $P=z\otimes\zd$,
namely
\be
\label{ELP} [\p\bp P, P]=0,
\ee

The necessary freedom is achieved if we use the homogeneous
unnormalised field variables $f$, such that
\be
\hspace{-1.5cm} z=f/(f^\dagger\c f)^{1/2}, \quad
\mathbb{C}\ni\xi\mapsto
f(\xi,\bar{\xi})=\left(f_0(\xi,\bar{\xi}),...,f_{N-1}(\xi,\bar{\xi})\right)\in\mathbb{C}^N\backslash\lbrace0\rbrace,
\ee
whose dynamics is governed by the unconstrained EL equations
\be
\label{E-L} \left(\mathbb{I}-\frac{f\otimes
f^{\dagger}}{f^{\dagger}\cdot f}\right)\cdot\left[\p\bp
f-\frac{1}{f^{\dagger}\cdot f}\left((f^{\dagger}\cdot\bp f)\p
f+(f^{\dagger}\cdot\p f)\bp f\right)\right]=0.
\ee
 The way in which these vector functions are constructed makes them
elements of a Grassmannian space Gr$(1, \mathbb{C}^N)$ \cite{Zak} and
suggests that equations (\ref{E-L}) are invariant under
multiplication of $f$ by any scalar function (which may easily be
checked by direct calculation). This property leaves too much
freedom for the shape of possible singularities. However if we
normalise the homogeneous variables in such a way that the first
component $f_0$ is equal to 1, we eventually obtain a system of
equation suitable for the Kovalevsky-Gambier analysis, commonly
known as the Painlev\'e test. The equations in terms of the affine
variables $w=(w_1,...,w_{N-1})$, such that
\be
w_i = f_i/f_0,~~~i=1,...,N-1\quad
\mathrm{(}\mathrm{generically}~f_0\ne 0\mathrm{)}.
\ee
read
\begin{subequations}
\bea
&&\left(1+\sum_{l=1}^{N-1}\bw_l w_l\right)\p\bp w_i - \sum_{l=1}^{N-1}(\bw_l\bp w_l\p w_i+\bw_l\p w_l\bp w_i)=0,\label{w-eqs-a}\\
&& \left(1+\sum_{l=1}^{N-1}w_l \bw_l\right)\bp\p w_i -
\sum_{l=1}^{N-1}(w_l\p \bw_l\bp \bw_i+w_l\bp \bw_l\p
\bw_i)=0,\label{w-eqs-b}
\eea
\end{subequations}
where the complex conjugates of (\ref{w-eqs-a}) have been written
separately as (\ref{w-eqs-b}) because the complex conjugation will
no longer link the variables $w_i$ with $\bw_i$ when we extend the
independent variables analytically to the double complex plane
$\mathbb{C}^2$ (as it is done in the Painlev\'e test). Therefore,
in what follows, we put quotation marks in `complex conjugation'
when we write about the symmetry which turns unbarred quantities into
the barred ones and vice versa.

Equations (\ref{w-eqs-a},~\ref{w-eqs-b}) will be the subject of
further analysis. They constitute a system of $2(N-1)$
second-order PDEs, which requires $4(N-1)$ first integrals to
build the general solution.

\section{The Painlev\'e test}

To perform the test, we look for the solution of the system
(\ref{w-eqs-a},~\ref{w-eqs-b}), extended to the double complex
plane $(\xi,\bxi)\in\C^2$, in the form of a Laurent series about a
movable noncharacteristic singularity manifold
\be\label{Kruskal}
\F(\xi, \bxi)= \bxi-\f(\xi)\quad \mathrm{(Kruskal's~
simplification)},
\ee
where the function $\f$ defining the singularity manifold is a
holomorphic function of $\xi$, while the coefficients of the
expansion are analytic in their arguments $(\xi,\bxi)$.

 The condition of being noncharacteristic excludes the
surfaces $\xi=0$ and $\bxi=0$, which in turn eliminates locally
holomorphic and locally antiholomorphic functions $w,~\bw$,
including the solutions of Din and Zakrzewski \cite{DZ,Din}. On
the other hand, the selection of non-characteristic singularity
manifolds makes possible both the Kruskal simplification
(\cite{JKM}) and the assumption $\f'(\xi)\ne 0$.

In the series below, we adopt the notation in which a superscript
for $\F$ is simply an exponent, while a superscript for a
dependent variable, e.g. $w_i^n$ denotes the $n$-th order
coefficient in the Laurent expansion of $w_i$. Additionally, it is
convenient to extend the notation to negative $n$, assuming
\be\label{convention}
w_i^n = 0\quad \mathrm{whenever}~ n<0,~ \mathrm{for~all~}
i=1,...,N-1.
\ee
We do not limit the number of dependent variables $w_i$ and allow
a priori the possibility that the initial exponents at each $w_i$
may be different. Thus the Laurent expansion has the form
\begin{subequations}\label{Laurent}
\bea\label{}
&&w_i=\sum_{n=0}^\infty w_i^n(\xi)\F^{n-\al_i},\\
&&\bw_i=\sum_{n=0}^\infty \bw_i^n(\xi)\F^{n-\beta_i},
\eea
\end{subequations}
where for all $i$ we have $w_i^0\ne 0$ and $\bw_i^0\ne 0$
(otherwise we would start from higher-order terms). We also assume
that $\al_i>0$ and $\beta_i>0$ for all $i$.

Let us substitute (\ref{Laurent}) into our equations
(\ref{w-eqs-a},~\ref{w-eqs-b}). As these equations are of 3rd
degree, the resulting equations contain quadruple sums (a sum over
the components of $w$ and products of 3 sums of the Laurent
series). We first rearrange the latter sums
$\sum_{m=0}^\infty\sum_{n=0}^\infty\sum_{p=0}^\infty~\to~\sum_{k=0}^\infty\sum_{n=0}^k\sum_{p=0}^n$,
where $k=m+n+p$. Then we shift the dummy index $k$ in such a way
that all terms indexed with the same $k$ become proportional to the same
powers of $\F$ (the range of $k$ remains unchanged thanks to our
convention (\ref{convention})). Next we require that the
coefficients of the same powers in $\F$ vanish.

Under the assumption that $\al_i>0,~\beta_i>0$ for all $i=0,...,N-1$,
there is no balance of terms with exponents of different form in
the lowest order. Therefore, the initial exponents are obtained
from the equations satisfied by the coefficients of the
lowest-order terms $k=0$ rather than those satisfied by the
exponents of those terms. In the lowest-order, we obtain for
$i=1,...,N-1$
\bs\label{zero order}
\bea
&&w_i^0\f'\al_i\sum_{l=1}^{N-1}\bw_l^0 w_l^0(2\al_l-\al_i-1) =
0\label{zero order-a}\\
&&\bw_i^0\f'\beta_i\sum_{l=1}^{N-1} w_l^0
\bw_l^0(2\beta_l-\beta_i-1) = 0,\label{zero order-b}
\eea
\es
where the prime denotes the derivative with respect to $\xi$ (we
have omitted the $\xi$-dependence of $\f$ and all the $w$'s).

Equations (\ref{zero order}), divided by $w_i^0\f' \al_i$ or
$\bw_i^0\f' \beta_i$, constitute two separate systems of linear
equations: one for $\al_1,...,\al_{N-1}$ and a similar one for
$\beta_1,...,\beta_{N-1}$. It is evident that
\be\label{exp-solution}
\al_1=,...,=\al_{N-1}=\beta_1=,...,=\beta_{N-1}=1
\ee
solves both systems (\ref{zero order}). A question arises as to
whether this is the only solution with all positive $\al_i$ and
$\beta_i$. We will show that this is indeed the case. The proof
below is performed for the system (\ref{zero order-a}). The proof
for (\ref{zero order-b}) is identical.

Proof. Let our movable singularity manifold intersects the plane
$\bxi=\xi^*$ (the plane of real $\xi^1$ and $\xi^2$), where the
asterisk denotes the actual complex conjugation. Consider the
matrix of coefficients, $B_{ij}=2\bw_i^0 w_i^0
-\delta_{ij}\sum_{l=1}^{N-1}\bw_l^0 w_l^0$. For each
$j=1,...,N-1$, all elements of the $j$-th column are identical, with
the exception of the diagonal element. If we add all the columns
to the first one and then subtract the first row of the resulting
matrix from each of the other rows, we obtain a triangular matrix,
with zeros everywhere except for the first row and the main
diagonal. The diagonal has $\sum_{l=1}^{N-1}\bw_l^0 w_l^0$ in the
first row and minus this sum in all other rows. Hence the
determinant of the system can be calculated explicitly
\be\label{det0}
\det\, B = -\left(\sum_{l=1}^{N-1}\bw_l^0 w_l^0\right)^{N-1}.
\ee
On the complex plane $\bxi=\xi^*$ the barred $w^0$'s are indeed
the complex conjugates of their unbarred counterparts. Hence all
the components of the sum in (\ref{det0}) are positive on this
plane, as we have assumed that $w_l^0$ and $\bw_l^0$ are nonzero
for all $l$. Continuity of these coefficients ensures that they
remain positive in some neighbourhood of this plane. Thus the
determinant (\ref{det0}) is nonzero (negative) and the solution
(\ref{exp-solution}) is unique in some domain. But the initial
exponents have to be independent of $(\xi,~\bxi)$, hence the
solution (\ref{exp-solution}) is unique in the neighbourhood of
the whole singularity manifold. \qed

Putting all $\al_i$ and $\beta_i$ equal to 1, as in
(\ref{exp-solution}), we obtain the recurrence relations at $k>0$.
For each $k=1,2,...$ they have the form of a system of $2(N-1)$
linear algebraic equations with the unknowns $w_1^k,...,w_{N-1}^k$
and $\bw_1^k,...,\bw_{N-1}^k$
\bea\label{recurrence-w}
&& (k-1)k\f'\left(\sum_{l=1}^{N-1}\bw_l^0 w_l^0 w_i^k+2k
\sum_{l=1}^{N-1}\bw_l^0 w_i^0 w_l^k\right)\nn\\
&&=-\f'\sum_{l=1}^{N-1}\sum_{n=1}^{k-1}\left[\sum_{p=0}^{k-n}
(n-1)(n-2p)w_i^0 \bw_l^{k-n-p} w_l^p+2n
w_i^0\bw_l^{k-n}w_l^n\right]\nn\\
&&+\sum_{l=1}^{N-1}\sum_{n=1}^{k-1}\left[\sum_{p=0}^{k-n}
\bw_l^{k-n-p-1}\left((n-p)w_l^p
(w_i^n)'+(n-1)(w_l^p)'w_i^n\right)\right]\nn\\&&+(k-3)(k-4)\f'
w_i^{k-2}-(k-4)(w_i^{k-3})',
\eea
and a similar set of $N-1$ equations for the `complex conjugates'
$\bw_1^k,...,\bw_{N-1}^k$. Note that the unknowns
$\bw_1^k,...,\bw_{N-1}^k$ are absent from the left-hand sides
(lhs) of (\ref{recurrence-w}) and similarly, the unknowns
$w_1^k,...,w_{N-1}^k$ are absent from the lhs of the conjugate
system (although the systems remain coupled with each other
through the right-hand sides (rhs)). This absence means that the
matrix of coefficients of the complete linear system is a direct
sum of two square matrices and its determinant is a product of
their determinants. The Fuchs indices or resonances (we use the
second name to avoid misunderstandings in our multi-index
notation) are calculated from the requirement that the determinant
vanish.

The first matrix has the elements
\be
A_{ij}^k = \f'
k\left[(k-1)\delta_{ij}\left(\sum_{l=1}^{N-1}\bw_l^0
w_l^0\right)-2 w_i^0\bw_j^0\right].
\ee
The second component of the direct sum is its `complex conjugate'.
The instances in which the determinant of their direct sum
vanishes are listed in Table 1.\vspace{2ex}

\noindent\begin{tabular}{|c|c|l|}
  \hline
  Resonance & Multiplicity & Linear dependence of the rows \\
  \hline
  $k=0$ & $2(N-1)$ &Each row identically vanishes. \\
  \hline
  $k=1$ & $2(N-2)$ & Row no. $i,~\mathrm{where}~i=2,...,N-1$,
  is equal to the 1st row \\&& multiplied by $w_i^0/w_1^0$;
  the second component of the direct \\&&sum is the `c.c.' of the first one.\\
  \hline
  $k=-1$ & 2 & If each row $i$, where $i,~~i=1,...,N-1$ is multiplied by \\&&$\bw_i^0/\bw_1^0$ and the products are added together,
  the result is \\&&zero. The second component of the direct sum
 is \\&&the `c.c.' of the first one.\\
  \hline
\end{tabular}\vspace{2ex}

 Altogether we have $4N-4$ zeros. This number is equal to the total
order of the system of PDEs. Hence there are no more resonances.

We now test the compatibility of the resonances by checking
whether the rhs of the equations (\ref{recurrence-w}) have the
same linear dependence between rows as their lhs

For $k=0$, all terms on the rhs contain $w$ with a negative
superscript, which according to our convention (\ref{convention})
means that they are equal zero. Hence the whole rhs is equal to zero,
as it should be. Consequently, it leaves room for $2(N-1)$
arbitrary functions of $\xi$ (first integrals).

For $k=1$, the rhs of the $i$-th equation, $i=1,...,N-1$, reduces
to $w_i^0\sum_{l=1}^{N-1}\bw_l^0 (w_l^0)'$, i.e. all rows are
proportional. For instance, we may take the first row and write
each of the rows no. $i=2,...,N-1$ as equal to the first row
multiplied by $w_i^0/w_1^0\,$. This satisfies the linear
dependence condition of Table 1. This leaves room for another
$2(N-2)$ arbitrary functions.

The above verification of compatibility cannot be performed for
negative zeros. One of the two zeros $k=-1$ corresponds to the
arbitrariness of $\f(\xi)$. The compatibility of the other zero at
$k=-1$ remains unknown.

The verified zeros allow us to introduce a total of $4N-6$
arbitrary functions of $\xi$. These are $w_i^0$ and $\bw_i^0$ for
$i=1,...,N-1$ and $w_i^1$ and $\bw_i^1$ for $i=2,...,N-1$.
Together with the arbitrary singularity manifold $\f$
(corresponding to one of the two zeros $k=-1$), they constitute a
set of $2N-5$ first integrals. There remains the second zero
$k=-1$, which is the cause of the missing $(4N-4)$-th first
integral. This problem will be addressed in the next section.

\section{The question of the double resonance at $k=-1$}
The negative resonances, except for a single $k=-1$ resonance,
correspond to essential singular points. In the $\cpn$ model, they
are connected with the coupling between $w$ and $\bw$ (if not for
the coupling we would have two separate systems, each possessing a
single resonance $k=-1$). A singularity connected with the phase
may indeed be essential. A question arises: does the essential
singularity introduce multivaluedness in the solution or not.

The authors tried the perturbative Painlev\'e analysis of
\cite{CFP} for the $\cp1$ model. Up to the third order in the
perturbation of the Laurent series (\ref{Laurent}) all the
resonances are compatible. However the order at which an
incompatibility may occur is difficult to predict. Being unable to
prove the Painlev\'e or non-Painlev\'e property by any systematic
method, we limit ourselves to a counterexample.

An example of a solution (an envelope solitary wave) which has
branching at a point dependent on the initial conditions has been
derived by Lie group analysis and the corresponding symmetry
reduction of the $\cp1$ model in \cite{GS1,GS2}. A typical
solution of the kind reads
\bea\label{example}
&&w(\xi,\bxi)=R \exp[i(\xi/a-f)],\quad \bw(\xi,\bxi)=R \exp[-i(\xi/a-f)],~~\mathrm{where}\nn\\
&&R = \pm\sqrt{\frac{(p-1) \cosh\, g\, +p+1}{(p-1) \cosh
\,g\,-p-1}},\nn\\
&&f
=\arctan\left(\frac{p+1}{2\sqrt{-p}}\tanh\,g\right)+\frac{\left(p+2\sqrt{-p}-1\right)\chi-2\sqrt{-p}\,\chi_0}{2(p-1)}+d,\nn\\
&&\mathrm{and}~~g = \frac{(p+1) \left(\chi-\chi_0\right)}{2
(p-1)}, \quad \mathrm{where}~~ \chi=\frac{\xi}{a}-\frac{\bxi}{b}.
\eea
To ensure that $w$ and $\bw$ are complex conjugates of each other
when the remaining quantities are real, it is usually assumed that
$p<-1$, however the solution is valid for any $p$.

 This solution (as well as several other solutions in the form
of elliptic functions) is associated with multileaf surfaces
\cite{GS1,GS2}. It is singular for $\chi-\chi_0=(k+1/2)
i\,\pi,~k\in\mathbb{Z}$. For these values of $\chi$, the argument
of $\arctan$ in (\ref{example}) becomes infinite, which results in
branching (i.e. multivaluedness of the $\arctan$ function). These
singularities do not lie on characteristics ($\xi=const$ and
$\bxi=const$), which makes them proper for the analysis. The
position of the singularities depends on four parameters:
$p,~a,~b,~\chi_0$, and thus also on the initial conditions, which
contradicts the usual understanding of the Painlev\'e property.
However the authors are aware that a more constructive answer to
the question of compatibility at the negative resonance would be
provided by a non-Painlev\'e solution with its position dependent
on an arbitrary function rather than a few parameters. We do not
have such a solution.

The action integral for the example (\ref{example}) is not finite,
hence it is compatible with the theorem of Din and Zakrzewski
\cite{DZ,Din,Zak}. Neither does it contradict the classical result
of \cite{ARS}, because it cannot be obtained as a solution of a
Gelfand-Levitan-Marchenko equation \cite{GLM} with a finite
integral kernel.

The $\cp1$ model is a limit case of $\cpn$ models, where all
but one affine coordinates (and all but one `complex conjugates')
tend to zero. Thus the absence of the Painlev\'e property in the
$\cp1$ infers its absence for all $\cpn$ models.

\section*{Conclusion}
We have shown that the equations governing the behaviour of $\cpn$
models, without the constraint of finite action, may have
solutions with movable singularities in the form of pole
manifolds. The order of the poles is one for all dependent
variables (the calculation based on the usual assumption of the
Painlev\'e test, i.e. negative initial exponents, eliminates poles
of other orders). For the $\cpn$ model equations, the Laurent
series about a pole manifold is consistent at all $4N-5$
nonnegative resonances. This way, it provides a family of
solutions with $4N-5$ parameter functions (first integrals) within
the domain of convergence of the series. However, branching may
still occur at essential singular points. We have provided an
example of a solution which is multivalued in the neighbourhood of
a sequence of non-characteristic movable singular manifolds. Their
position depends on the initial conditions through four
parameters. It would be desirable to find a deformation of such
solutions turning them into solutions depending on an arbitrary
function.

The Painlev\'e analysis is nontrivial for these models due to the
indefinite number of equations and dependent variables.

\ack AMG's work was supported by a research grant from NSERC of
Canada. P.P.G. wishes to thank the Centre de Recherches
Math\'ematiques (Universit\'e de Montr\'eal) for the NSERC
financial support provided for his visit to the CRM.


\section*{References}
\begin{thebibliography}{99}


\bibitem{ARS}      Ablowitz M J, Ramani A and Segur H (1980) A connection between nonlinear evolution equations and ordinary differential equations of P-type \textit{J. Math. Phys.} \textbf{21} 715--721 and 1006--1015.
\bibitem{Bab1}  Babelon O (2007)  A short introduction to Classical and Quantum Integrable Systems, Univ. Paris 6.
\bibitem{Bab2}  Babelon O, Bernard D and Talon M (2003)  Introduction to Classical Integrable Systems, Cambridge University Press.
\bibitem{Carrol}    Carrol R and Konopelchenko B (1996) Generalized Weierstrass-Enneper inducing conformal immersions and gravity \textit{Int. J. Mod. Phys. A} \textbf{11} 1183--1216.
\bibitem{Conte}     Conte R (1999) The Painlev\'e approach to nonlinear ordinary
differential equations, chapter 3, 77--180 in The Painlev\'e
Property One Century Later, ed. R. Conte, New York, Springer
Verlag.
\bibitem{CM}    Conte R and Musette M (2008) The Painlev\'e Handbook, Dordrecht,
Springer.
\bibitem{CFP}   Conte~R, ~Fordy~A.~P  and ~Pickering~A (1993),
A perturbative Painlev\'e approach to nonlinear differential
equations, \textit{Physica D} {\bf 69}  33--58.
\bibitem{Dav}   Davydov A (1999) Solitons in Molecular Systems, New York Kluver.
\bibitem{Din}   Din A M (1984) The Riemann-Hilbert problem and finite-action CPN?1 model solutions, \textit{Nucl Phys B} 233,269--288.
\bibitem{DHZ}
Din A.M., Horvath Z. and Zakrzewski W.J. (1984) The
Riemann--Hilbert problem and finite action $\mathbb{C}P^{N-1}$
solutions, \textit{Nucl. Phys. B} \textbf{233} 269.
\bibitem{DZ}    Din A M, Zakrzewski W J, (1980) General classical solutions in the $\cpn$ model \textit{Nucl Phys B} \textbf{174} 397--406.
\bibitem{Gross} Gross D G, Piran T and Weinberg S (1992) Two-dimensional Quantum Gravity and Random Surfaces, Singapore: World
Scientific.
\bibitem{GS1}   Grundland A M and \v{S}nobl (2006) Description of surfaces associated with $\cpn$ sigma models on Minkowski space \textit{J Geom Phys} \textbf{56} 512--531.
\bibitem{GS2}   Grundland A M and \v{S}nobl (2006) Surfaces Associated with Sigma Models  \textit{Stud Appl Math} \textbf{117} 335-351.
\bibitem {JKM}  Jimbo M, ~Kruskal~M~D and ~Miwa~T (1982) Painlev\'e test for the self-dual Yang-Mills equation, \textit{Phys.~Lett.~A} {\bf 92}  59--60.
\bibitem{Landolfi}  Landolfi G (2003) On the Canham-Helfrich membrane model \textit{J Phys A: Math Theor} 36 4699 (16pp).
\bibitem{Manton}    Manton N.and Sutcliffe P (2004) Topological Solitons, Cambridge University Press
\bibitem{GLM}   Marchenko V A (2011) Sturm-Liouville operators and Applications AMS
\bibitem{Mus}   Musette M (1999) The Painlev\'e analysis for nonlinear partial
differential equations, chapter 8, 517--562 in The Painlev\'e
Property One Century Later, ed. R. Conte, New York, Springer
Verlag.
\bibitem{Raj}   Rajaraman R (2002) $CP_N$ Solitons in quantum Hall systems, \textit{Europhys J B} \textbf{28} 157-162.
\bibitem{WTC}   Weiss J, Tabor M and Carnevale G (1983) The Painlevé property for
partial differential equations, \textit{J. Math. Phys.} {\bf 24}
522--526.
\bibitem{Zak}   Zakrzewski W J (1989) Low Dimensional Sigma Models, ch. 4 and 8--11, Bristol, Adam
Hilger.

\end{thebibliography}
\end{document}